\documentclass[a4paper,11pt]{article}
\usepackage{pos}

\title{Development and performance of a compact LumiCal prototype calorimeter for future linear collider experiments}
 \ShortTitle{Development and performance of compact LumiCal}

\author*[a,1]{Maryna Borysova}
\note{For the FCAL collaboration}

\affiliation[a]{Institute for Nuclear Research,\\
  47, prospekt Nauky, Kiev, Ukraine}

\emailAdd{maryna.borysova@desy.de}

\abstract{The FCAL collaboration is preparing large-scale prototypes of special calorimeters to be used in the very forward region at future electron-positron colliders for a precise measurement of integrated luminosity and for instant luminosity measurement and assisting beam-tuning. LumiCal is designed as a silicon-tungsten sandwich calorimeter with very thin sensor planes to keep the Molière radius small, facilitating such the measurement of electron showers in the presence of background. Dedicated front-end electronics has been developed to match the timing and dynamic range requirements. 
A partially instrumented prototype was investigated in a 1 to 5 GeV electron beam at the DESY II synchrotron. 
In the recent beam tests, a multi-plane compact prototype was equipped with thin detector planes fully assembled with readout electronics and installed in 1 mm gaps between tungsten plates of one radiation length thickness. High statistics data were used to perform sensor alignment, and to measure the longitudinal and transversal shower development in the sandwich. 
This talk covers the latest status of the calorimeter prototype development and selected performance results, obtained in test beam measurements, the prospects for the upcoming DESY test beam, as well as the expected simulation performance.}

\FullConference{%
  40th International Conference on High Energy physics - ICHEP2020\\
  July 28 - August 6, 2020\\
  Prague, Czech Republic (virtual meeting)
}

\begin{document}
\maketitle

\section{Introduction}
Forward calorimeters for future electron-positron linear collider experiments have challenging requirements on geometrical compactness and high precision measurements of integrated luminosity~\cite{FCAL_ILC} resulting in the design of highly compact calorimeters. The FCAL Collaboration is designing two such calorimeters that would cover the far-forward region of proposed future electron-positron linear colliders~\cite{ILC_TDR_v4_det,ILC_TDR_v1_phys,CLIC_UPDATE_YP}. Precise measurement of integrated luminosity is provided by the LumiCal detector. Another detector, BeamCal, is designed for instant luminosity measurement and assisting beam tuning when included in a fast feedback system. Both detectors extend the capabilities of the experiments for physics study in the high rapidity region.

Luminosity in  e$^{+}$e$^{-}$ collider experiments can be measured using Bhabha scattering, as a gauge process (e$^{+}$e$^{-}\rightarrow$~e$^{+}$e$^{-}$($\gamma$)).
In this case the luminosity is obtained as a ratio of the number of Bhabha events in a certain polar angle range to the integral of differential cross section 
$L = N_{B} / \sigma_{B}$. 
The cross section can be accurately calculated in QED~\cite{Bhabha_scatt}. The range in the polar angle for the LumiCal baseline design was optimised in a simulation~\cite{FCAL_ILC}. The statistical uncertainty of the luminosity measurement is defined by the number of Bhabha 
events identified by LumiCal. The fiducial volume in a polar angle range considered is the part of the polar angle range where electron showers are fully contained. Keeping it as large as possible allows reducing the statistical uncertainty of luminosity measurement. This is achieved by designing a calorimeter with a small Molière radius. A small Molière radius is also important for BeamCal to improve high energy electrons reconstruction on top of background.  

\section{Compact LumiCal prototype}


 \begin{figure}[h!]
  \begin{minipage}[t]{0.33\textwidth}
      \includegraphics[width=\textwidth]{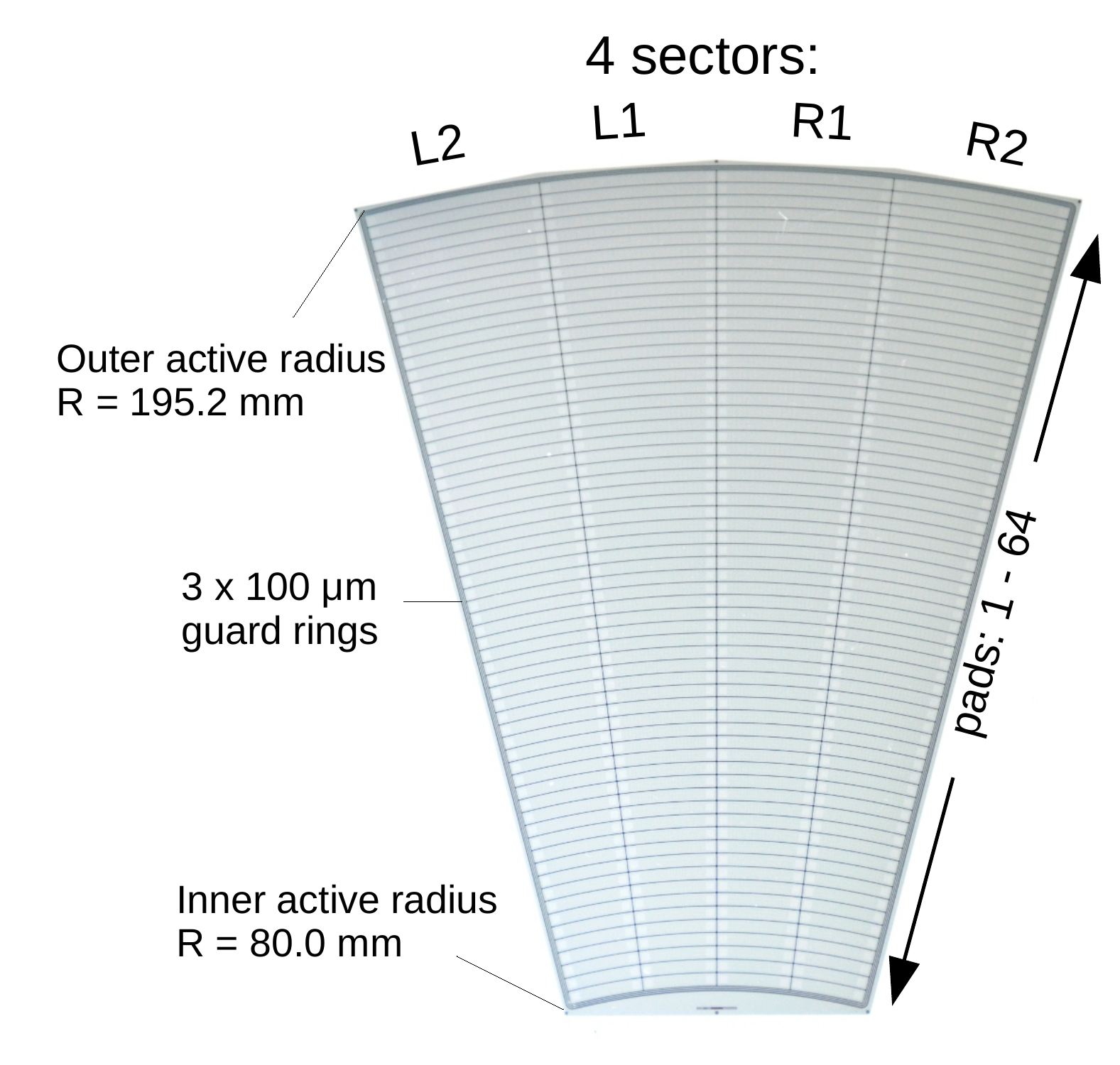}
    \caption{The LumiCal silicon sensor prototype. }
    \label{LumiCalsensor}
  \end{minipage}\hfill
  \hspace{0.025\textwidth}
  \begin{minipage}[t]{0.60\textwidth}
      \includegraphics[width=\textwidth]{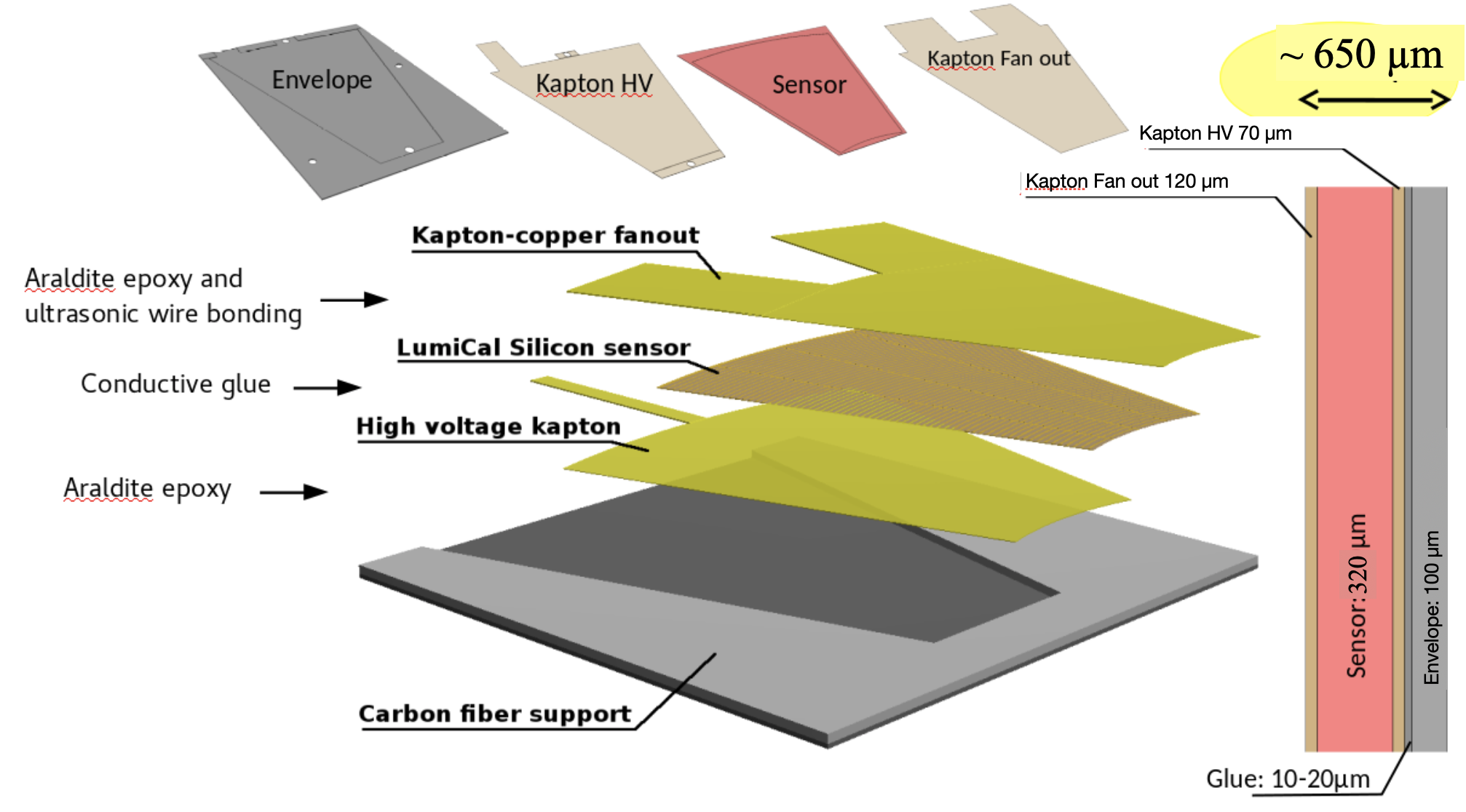}
  \caption{The LumiCal detector module assembly. }
    \label{assembly}
  \end{minipage}
\end{figure}

The design of a LumiCal sensor was optimized in simulations to provide the required resolution of the polar angle reconstruction. Tungsten absorber plates of one radiation lenght thickness are interspersed with silicon detector planes.
A silicon sensor prototype is shown in Fig.\ref{LumiCalsensor}. It is shaped as a ring segment of 30 degrees, subdivided into four sectors of 7.5 degrees each. There are 64 radial p-type pads in each sector with a pitch of 1.8 mm. The sensor is about 11 cm long with an inner radius of 80 mm. It is made of  n-type silicon  with the thickness of  320~$\mu m$. The compactness of the calorimeter is achieved by a thin detector module design as shown in Fig.\ref{assembly}.
In the current prototype the space between tungsten plates is 1 mm. Fig.~\ref{assembly} shows a detector module of about 650~$\mu m$ thickness to be installed in the 1 mm gap.
Two thin Kapton PCBs are used to supply high voltage and to connect read-out electronics to the sensor. The first one is connected by conductive glue and the second by ultrasonic wire bonding. 
A carbon fiber support provides mechanical stiffness and facilitates handling and mounting between tungsten plates. 

To allow the insertion of the detector planes without damage, exceptionally flat tungsten plates are needed.
To facilitate machining, tungsten plates are made of an alloy with 93~\% tungsten, 5~\% nickel and 2~\% copper. The measurements showed that the flatness of the tungsten plates is better than 30~$\mu m$.
They are glued to permaglass frames that accurately determine their positions in the calorimeter stack.

\section{The dedicated readout ASIC}
\begin{figure}[htbp]
  \begin{center}
    \includegraphics[width=0.65\textwidth]{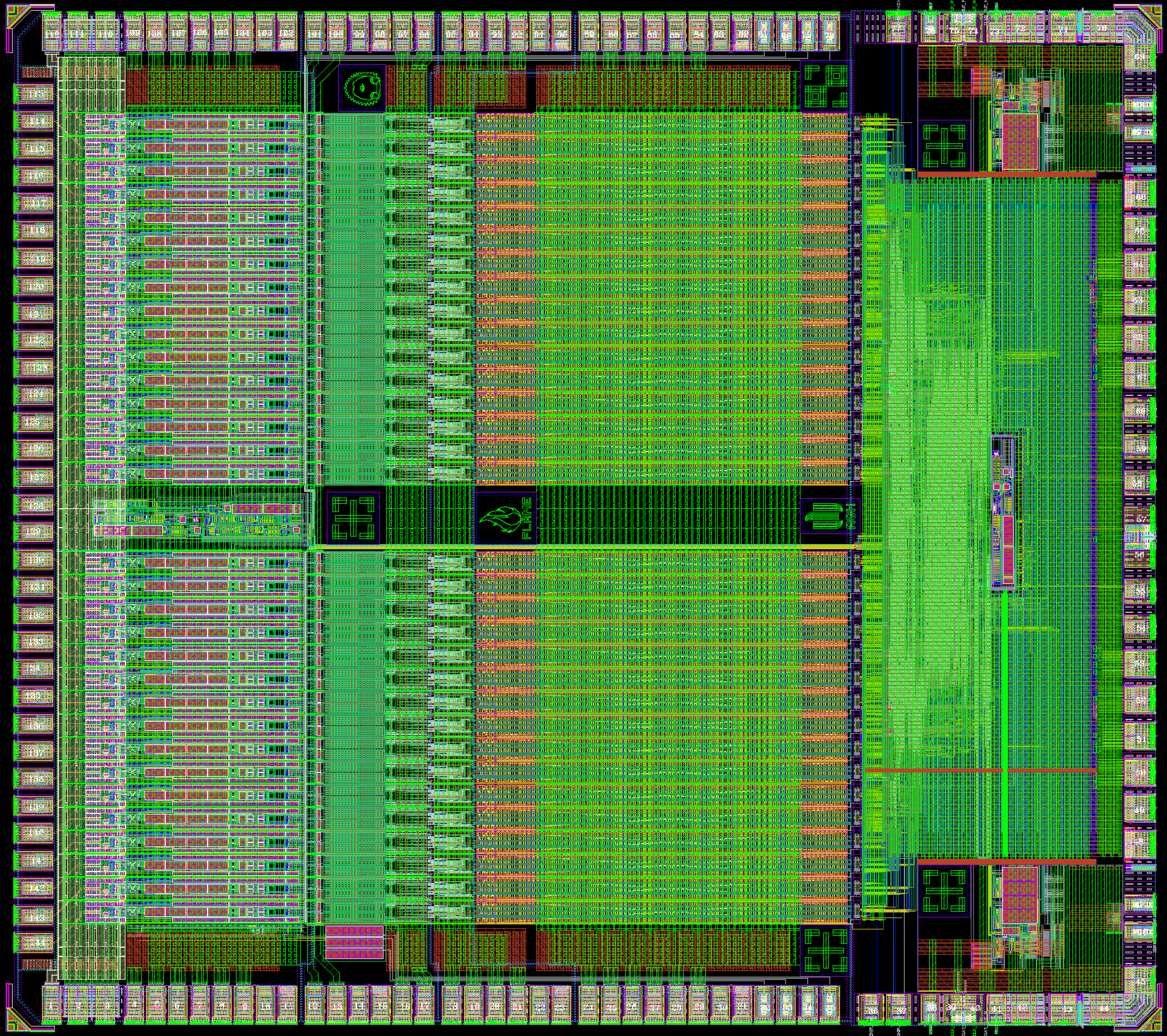}
      \end{center}
    \caption{The FLAME layout, the size is 3.7mm x 4.3mm }
    \label{fig:flame}
\end{figure}

A dedicated readout ASIC for LumiCal - FLAME - is being developed. 
This is a complete readout ASIC with a full functionality, including front-end readout, ADC, biasing and calibration fabricated in 130~nm CMOS technology. The FLAME layout with dimensions of 3.7~mm x 4.3~mm is shown in Fig.~\ref{fig:flame}. 
 It has 32 mix-mode channels. Each channel comprises variable gain front-end; 10 bit ADC with sampling rate up to 50 mega samples per second with power consumption, below 2~mW/channel. 
FLAME also contains a fast serializer with transmission rate up to 8 Gbps.

\section{The performance tests}
\begin{figure}[h!]
  \centering
 \includegraphics[width=0.9\textwidth]{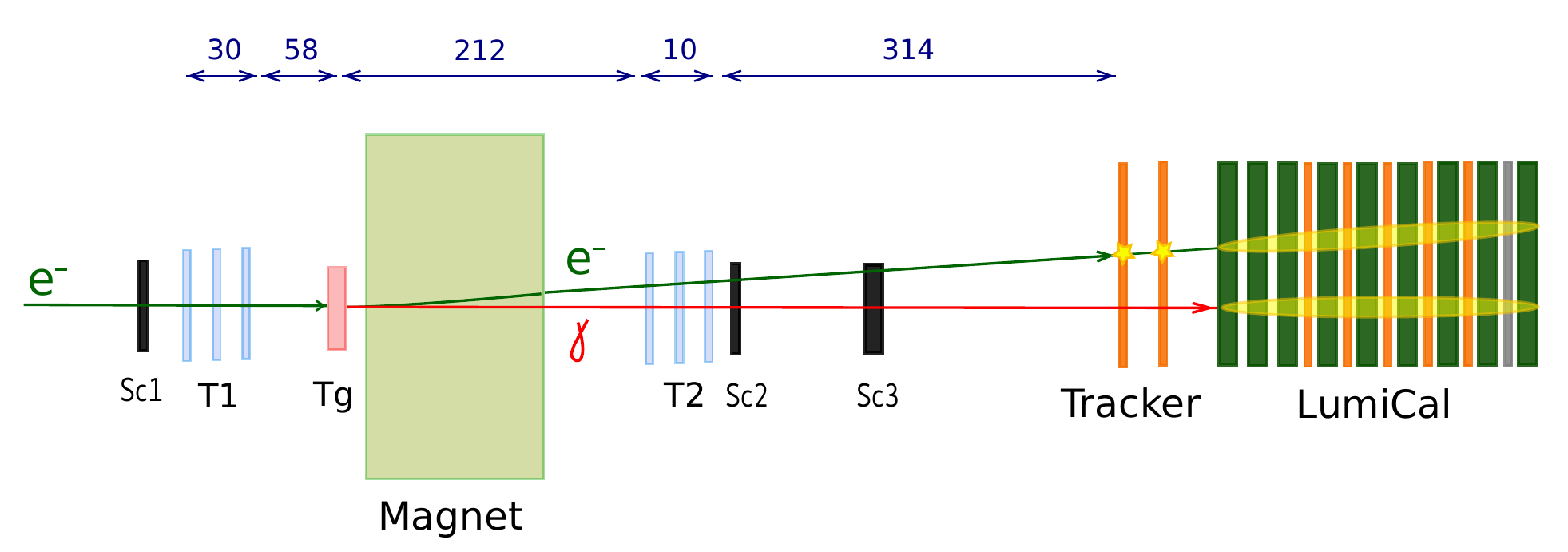}
  \caption{Geometry of the beam test setup (not to scale). Sc1, Sc2 and Sc3 are scintillator counters; 
           T1 and T2~ the arms of three-planes pixel telescope, Tg~the copper target for bremsstrahlung photon production and LumiCal, the calorimeter prototype under test. Distances, rounded to integer numbers in centimetres, are shown in the upper part of the figure.}
  \label{tb2016}
\end{figure}

Measurements were perfomed  in two beam test campaigns in 2016 and 2020 at the DESY-II Synchrotron using electrons with energies between 1~GeV and 5~GeV. A sketch of the experimental setup is shown in Fig.~\ref{tb2016}.
The main goal of the 2016 test beam was to study the performance of the ultra compact design.  

\begin{figure}[h!]
  \begin{minipage}[t]{0.45\textwidth}
      \includegraphics[width=\textwidth]{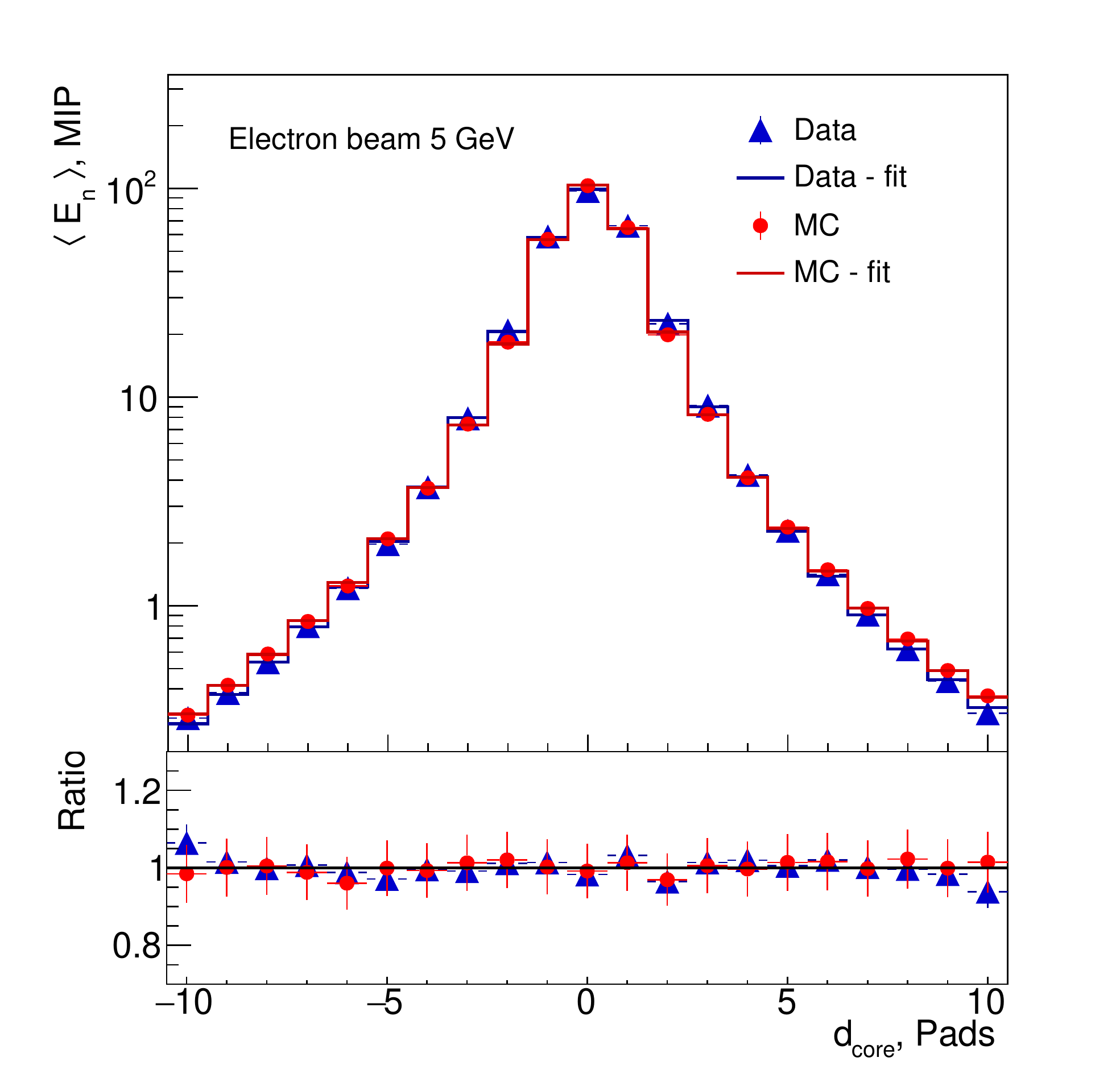}

    \caption{The average transverse shower profile as a function of the distance from the core, $d_{core}$, in units of the pad dimension (1.8~mm), for data (blue triangles) and MC simulation (red circles). The distributions are obtained with a 5 GeV electron beam.}
    \label{MR_5GeV}
  \end{minipage}\hfill
  \hspace{0.025\textwidth}
  \begin{minipage}[t]{0.45\textwidth}
      \includegraphics[width=\textwidth]{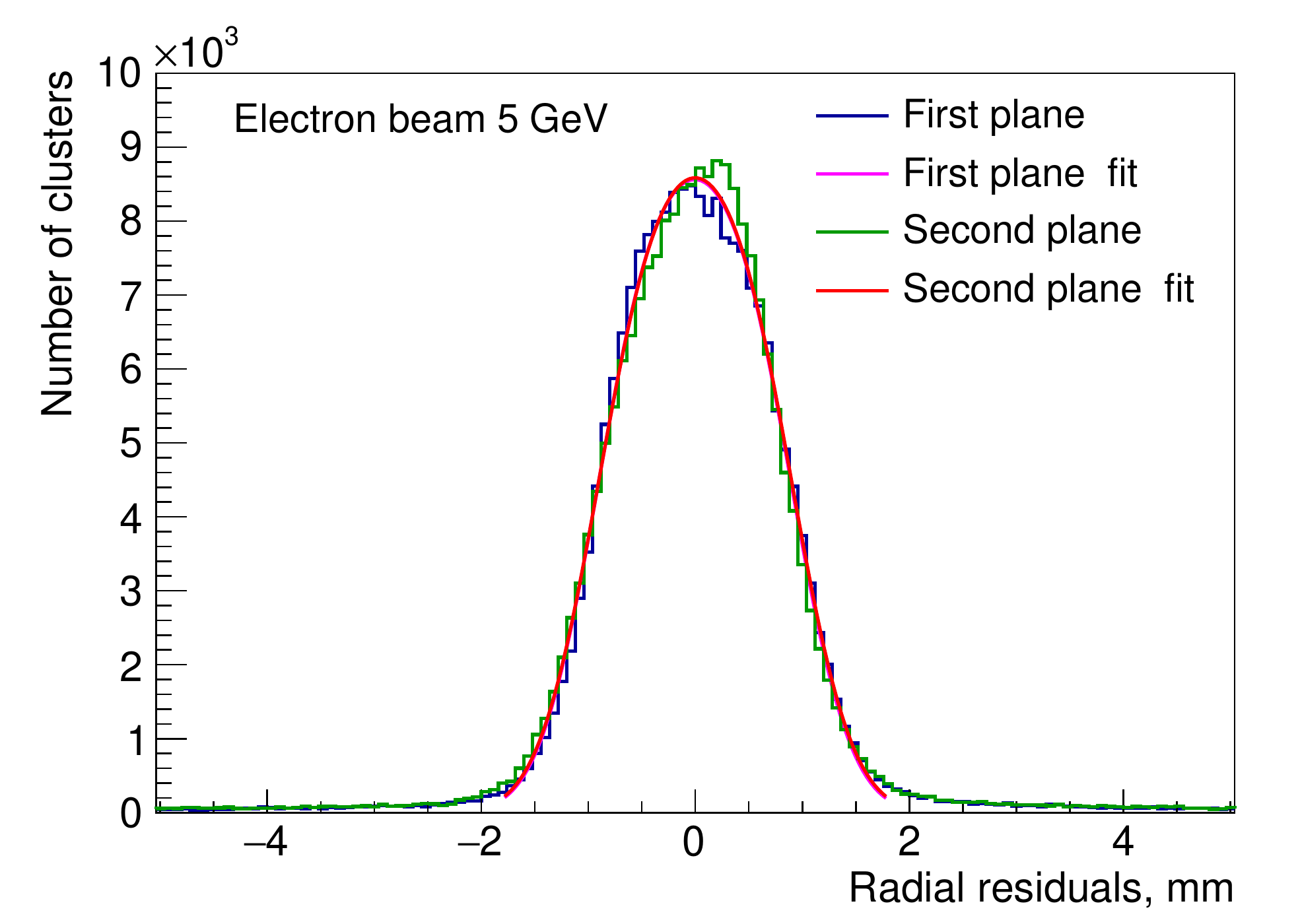}
  \caption{Distribution of residuals of the radial position measurements in the tracking planes and the calorimeter, obtained with a 5 GeV electron beam. }
    \label{residuals}
  \end{minipage}
\end{figure}

The average transverse shower profile as a function of the distance from the core, in units of the pad dimension (1.8~mm) is shown in Fig.~\ref{MR_5GeV}.
The lower part of the figure shows the ratio of the distributions to the fitted function, for the data (blue) and simulation (red). 
The segmentation of the LumiCal sensor is rather specific and a method was developed~\cite{LumiCal_multilayer_tb2014_epjc} to take into account its circular shape to measure the Molière radius. The effective Molière radius is found to be: (8.1 +/- 0.1(stat.) +/-0.3(syst.)) mm~\cite{LumiCal_multilayer_tb2019_epjc}.
 
The resolution of the shower position reconstruction in the calorimeter is estimated using the tracker planes in front of the calorimeter  denoted as `Tracker' in Fig.~\ref{tb2016}. 
Fig.~\ref{residuals} shows the distribution of the residuals of the reconstructed radial position of the shower in the calorimeter and in the two planes of the tracker. The position resolution amounts to (440+/-20)~$\mu m$ for 5 GeV electrons. It is in agreement with the simulation and the design for ILC energies.

\begin{figure}[htbp]
  \begin{center}
    \includegraphics[width=0.75\textwidth]{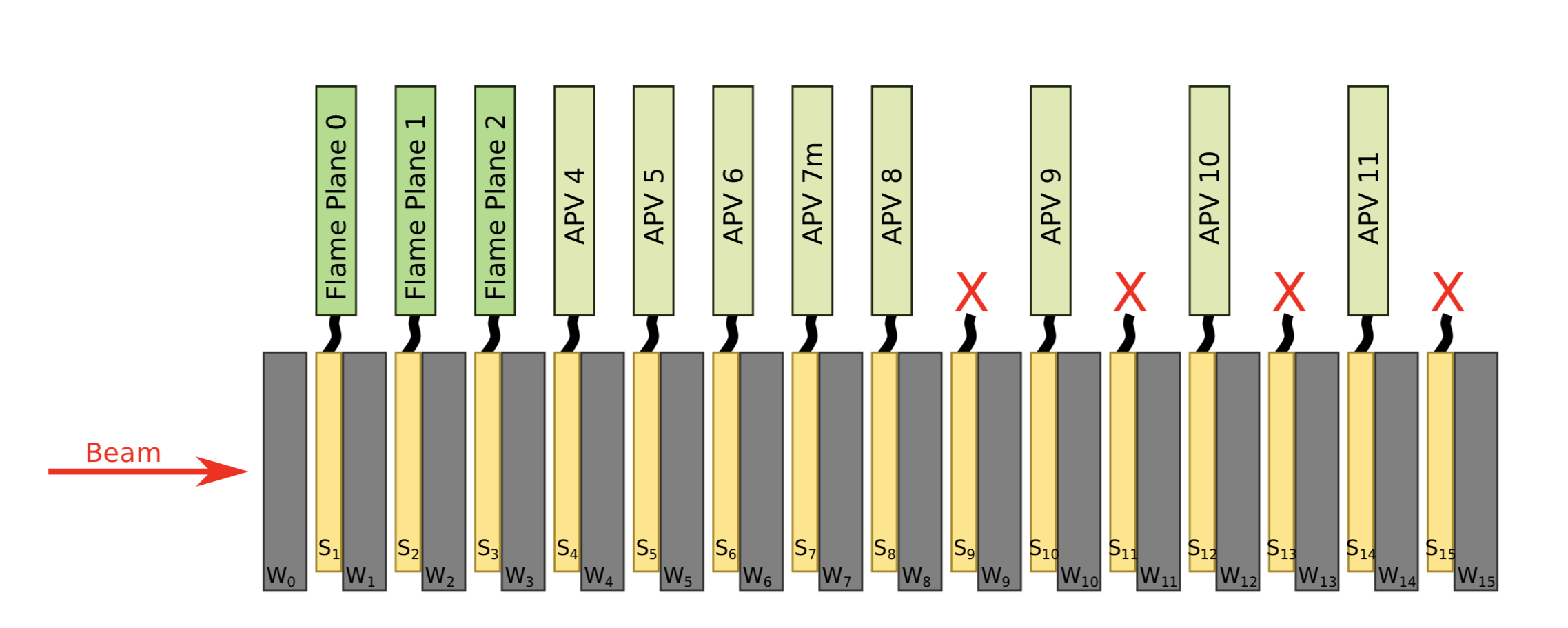}
      \end{center}
    \caption{The sketch of experiment layout }
    \label{fig:sketch}
\end{figure}
In addition, the data taken for an electron beam of 5 GeV are used to measure the efficiency of electron/gamma identification using two tracking planes in front of the calorimeter. The geometry of the setup for this study is optimized in a simulation. Bremsstrahlung photons are produced by the electron beam hitting the copper target installed upstream close to the dipole magnet. The magnetic field is chosen to allow both photons and electrons to travel within the acceptance of the second telescope T2 and arrive to LumiCal at a distance between them large enough to be resolved in the calorimeter. The efficiency of electron/gamma identification is estimated to be more than 90\% at 2.5 mm matching distance.

The most recent test beam 2020 campaign was dedicated to studies of the performance of deep LumiCal prototype with 15 sensitive layers.
In these tests, 3 planes were equipped with FLAME and the others - with double-gain readout using APV25 as seen in Fig.~\ref{fig:sketch}.
In addition, edge-scans for a fiducial volume study were carried out, electron/gamma response was tested and 
data with a tilted calorimeter were collected to study the reconstruction of showers in ILC-like geometry. The data analysis is in progress.

\section{Summary}
FCAL has developed a design for the very forward region for experiments at e$^{+}$e$^{-}$  colliders. The LumiCal calorimeter is foreseen for the precise measurement of the integrated luminosity  and 
BeamCal - for bunch-by-bunch luminosity measurement and high-energy electron tagging.
Silicon sensors for prototype of LumiCal are designed and fabricated.
Dedicated front-end ASICs are designed and fabricated in 130~nm CMOS technology. Prototypes of fully instrumented detector planes are built and tested.
A prototype of a highly compact calorimeter was studied in test-beams at DESY
resulting in an effective Molière radius of 8~mm at 5 GeV and
the shower position reconstruction of 440~$\mu m$ resolution at 5 GeV. The simulations are in agreement with the data.
Technologies developed in FCAL are applied in other experiments, e.g. CMS, XFEL and considered for LUXE at DESY.

\acknowledgments
This study was partly supported by the Israel Science Foundation (ISF), 
Israel German Foundation (GIF), the I-CORE program of the Israel Planning 
and Budgeting Committee, Israel Academy of Sciences and Humanities, by the 
National Commission for Scientific and Technological Research (CONICYT - 
Chile) under grant FONDECYT1170345, by the  Polish  Ministry of 
Science and Higher Education under contract nrs 3585/H2020/2016/2 
and 3501/H2020/2016/2, the Romanian UEFISCDI agency under 18PCCDI/2018 project and grant no. 16N/2020, by the 
Ministry of Education, Science and Technological Development of 
the Republic of Serbia within the project Ol171012, by the United 
States Department of Energy, grant de-sc0010107, and by the 
European Union Horizon 2020 Research and Innovation programme 
under Grant Agreement no.654168 (AIDA-2020). The measurements leading to 
these results have been performed at the Test Beam Facility at DESY Hamburg 
(Germany), a member of the Helmholtz Association (HGF).


\begin{thebibliography}{99}
   \bibitem{FCAL_ILC}
   H. Abramowicz et al., \emph{Forward instrumentation for ILC detectors.} JINST {\bf 5} (2010) P12002 [1009.2433].
     \bibitem{ILC_TDR_v4_det}
   T.~Behnke et al.,
   \emph{The International Linear Collider. Technical Design Report, Volume 4}: Detectors, 2013. arXiv:1306.6329 [physics.ins-det].

   \bibitem{ILC_TDR_v1_phys}
   H.~Baer et al., \emph{The International Linear Collider. Technical Design Report, Volume 2: Physics}. 2013. arXiv:1306.6352 [hep-ph].
   
   \bibitem{CLIC_UPDATE_YP}
   The CLIC, CLICdp collaborations, \emph{Updated baseline for a staged Compact Linear Collider.} CERN Yellow Report CERN-2016-004;
    arXiv:1608.07537 [physics.acc-ph]. 
   
   \bibitem{Bhabha_scatt}
    D.~Bardin et al., \emph{One-loop electroweak radiative 
    corrections to polarized Bhabha scattering.} arXiv:1801.00125 [hep-ph].
    
    \bibitem{LumiCal_multilayer_tb2014_epjc}
    H. Abramowicz et al., \emph{Measurement of shower development and its Moli\`ere radius with a four-plane LumiCal test set-up.}
    Eur. Phys. J. {\bf C 78} (2018) 135 [1705.03885]. https://doi.org/10.1140/epjc/s10052-018-5611-9 
      
      \bibitem{LumiCal_multilayer_tb2019_epjc}
    H. Abramowicz et al., \emph{Performance and Molière radius measurements using a compact prototype of LumiCal in an electron test beam.}
    Eur. Phys. J. {\bf C 79} (2019) 579 [1812.11426]. http://dx.doi.org/10.1140/epjc/s10052-019-7077-9 


\end{thebibliography}
\end{document}